\documentclass[aps,prl,twocolumn,showpacs,amsmath,amssymb, nofootinbib]{revtex4-1}
\usepackage{epsfig}
\usepackage{graphicx}
\usepackage{dcolumn}
\usepackage{bm}
\usepackage{ltablex,booktabs}
\usepackage{overpic}
\usepackage{subfigure}
\usepackage{float}
\usepackage{color}
\usepackage{amsmath}
\usepackage{mathcomp}
\usepackage{mathrsfs}
\usepackage{multirow}
\usepackage{rotating}
\usepackage{amssymb}
\usepackage{gensymb}
\usepackage{amsmath}
\usepackage{tabularx}
\usepackage{lineno}
\usepackage{tikz}
\usepackage[compat=1.1.0]{tikz-feynman}
\tikzfeynmanset{warn luatex=false}
\usepackage[bookmarksnumbered, pdfstartview=FitH,colorlinks,urlcolor=blue, citecolor=blue,linkcolor=blue] {hyperref}

\begin{document}
\normalsize
\parskip=5pt plus 1pt minus 1pt


\title{
Study of the light scalar $a_{0}(980)$ through the decay $D^{0} \to a_{0}(980)^-e^{+} \nu_{e}$ with $a_{0}(980)^- \to \eta \pi^-$
}
\author{
\begin{small}
\begin{center}
M.~Ablikim$^{1}$, M.~N.~Achasov$^{4,c}$, P.~Adlarson$^{76}$, O.~Afedulidis$^{3}$, X.~C.~Ai$^{81}$, R.~Aliberti$^{35}$, A.~Amoroso$^{75A,75C}$, Q.~An$^{72,58,a}$, Y.~Bai$^{57}$, O.~Bakina$^{36}$
% ...
\\
\vspace{0.2cm}
(BESIII Collaboration)\\
\vspace{0.2cm} {\it
$^{1}$ Institute of High Energy Physics, Beijing 100049, People's Republic of China\\
% ...
$^{3}$ Bochum  Ruhr-University, D-44780 Bochum, Germany\\
$^{4}$ Budker Institute of Nuclear Physics SB RAS (BINP), Novosibirsk 630090, Russia\\
% ...
$^{35}$ Johannes Gutenberg University of Mainz, Johann-Joachim-Becher-Weg 45, D-55099 Mainz, Germany\\
$^{36}$ Joint Institute for Nuclear Research, 141980 Dubna, Moscow region, Russia\\
% ...
$^{57}$ Southeast University, Nanjing 211100, People's Republic of China\\
$^{58}$ State Key Laboratory of Particle Detection and Electronics, Beijing 100049, Hefei 230026, People's Republic of China\\
% ...
$^{72}$ University of Science and Technology of China, Hefei 230026, People's Republic of China\\
% ...
$^{75}$ University of Turin and INFN, (A)University of Turin, I-10125, Turin, Italy; (B)University of Eastern Piedmont, I-15121, Alessandria, Italy; (C)INFN, I-10125, Turin, Italy\\
$^{76}$ Uppsala University, Box 516, SE-75120 Uppsala, Sweden\\
% ...
$^{81}$ Zhengzhou University, Zhengzhou 450001, People's Republic of China\\
\vspace{0.2cm}
$^{a}$ Deceased\\
% ...
$^{c}$ Also at the Novosibirsk State University, Novosibirsk, 630090, Russia\\
% ...
}
\end{center}
\vspace{0.4cm}
\end{small}
}
\noaffiliation{}

\date{\today}
  
\begin{abstract}
Using 7.93~${\rm fb^{-1}}$ of $e^+e^-$ collision data collected at a center-of-mass energy of 3.773~${\rm GeV}$ with the BESIII detector, we present an analysis of the decay $D^{0} \to \eta \pi^- e^+ \nu_{e}$.
The branching fraction of the decay $D^{0} \to a_{0}(980)^{-} e^+ \nu_{e}$ with $a_{0}(980)^{-} \to \eta \pi^{-}$ is measured to be $(0.86\pm0.17_{\text{stat}}\pm0.05_{\text{syst}})\times 10^{-4}$.
The decay dynamics of this process is studied with a single-pole parameterization of the hadronic form factor and the Flatt\'e formula describing the $a_0(980)$ line shape in the differential decay rate.
The product of the form factor $f^{ a_0}_{+}(0)$ and the Cabibbo-Kobayashi-Maskawa matrix element $|V_{cd}|$ is determined for the first time with the result  $f^{ a_0}_+(0)|V_{cd}|=0.126\pm0.013_{\rm stat}\pm0.003_{\rm syst}$.

\end{abstract}
\maketitle

Understanding the nature of scalar mesons has emerged as a pivotal challenge within the realm of non-perturbative quantum chromodynamics (QCD), given their important role in dynamically generating mass via the spontaneous breaking of QCD's chiral symmetry. Consequently, a thorough understanding of the phenomenon of confinement physics necessitates a deep exploration of the nature of these mesons~\cite{Achasov:2018grq, Jaffe:1976ig}. 
%

Although the existence of scalar mesons~($S_0$) below 1~GeV, including the $K_{0}^{*}(700)$, $f_0(500)$~($\sigma$), $f_0(980)$~($f_0$), and $a_{0}(980)$~($a_0$), has been firmly established, their identification within the constituent quark model~\cite{PDG} remains a long-standing puzzle, particularly due to the challenges encountered in describing them as normal quark-antiquark~($q\bar{q}$) states.
The two-quark model expects that the non-strange-flavored $a_0(980)$ should be as light as the $\sigma$ meson, whereas experimentally it is mixed with the strange-flavored $f_0(980)$.
Several experimental observations have notably intensified the discrepancies with the conventional $q\bar{q}$ interpretation.
The decay $\phi \to \gamma a_{0}(980)^0$, which violates both charge conjugation and parity conservation, is also observed with an anomalously large branching fraction~(BF)~\cite{PDG,Braghin:2022uih}.
In addition, the BESIII collaboration measured the BFs of the $D_s^+\to a_0(980)^{0(+)}\pi^{+(0)}$~\cite{BESIII:2019jjr}, $D_s^+\to a_0(980)^{0(+)}\rho^{+(0)}$~\cite{BESIII:2021aza} and $D^{0(+)}\to a_0(980)^{+}\pi^{-(0)}$\cite{BESIII:2024tpv} decays to be much larger than the expectations based on the naive two-quark model~\cite{Hsiao:2019ait, Yu:2021euw,Duan:2020vye}.

Various structure hypotheses have been proposed to explain these intriguing phenomena of scalar particles, including compact tetraquark states~($q^2\Bar{q}^2$)~\cite{Jaffe:1976ig,Brito:2004tv,Klempt:2007cp,Alexandrou:2017itd,Humanic:2022hpq}, two-meson molecule bound states~\cite{Weinstein:1982gc,Dai:2014lza,Sekihara:2014qxa}, and mixed states~\cite{Braghin:2022uih}. 
Nevertheless, the intricate interplay of non-perturbative strong interactions, combined with the scalar mesons sharing the same spin-parity quantum numbers as the QCD vacuum, leads to significant hadronic uncertainties in their classification.

Semileptonic~(SL) $D$ decays offers a laboratory where the mechanism of the $D \to S_{0}$ transition can be cleanly studied, given that the colorless lepton pair is not sensitive to the strong interaction~\cite{Ke:2023qzc, Ivanov:2019nqd, Hsiao:2023qtk, Wang:2009azc, CLEO:2009ugx, BESIII:2023wgr}. The form factor~(FF) of the $D \to S_{0}$ transition 
can be measured to provide a novel insight into the non-perturbative QCD effects on hadronization.
Moreover, under the constraints of SU(3) flavor symmetry, the variation in the value of the $f_{0}$-$\sigma$ mixing angle that is expected under different models leads to differences in the decay properties of $D \to S_{0} e^+ \nu_{e}$ with the BF and FF of the $D \to a_{0}(980) e^+ \nu_{e}$ as either a normalization or essential inputs~\cite{ Hsiao:2023qtk, Wang:2009azc}.
Consequently, the $D \to a_{0}(980) e^+ \nu_{e}$ decay not only offers an ideal probe into the nature of $a_{0}(980)$, but also gives the opportunity to distinguish between different descriptions of light scalar mesons.




BESIII collaboration reported the observation of $D^{0} \to a_{0}(980)^{-} e^+ \nu_{e}$ with a  25\% uncertainty in the measurement of  its BF~\cite{BESIII:2018sjg, Ke:2023qzc} using 2.93~fb$^{-1}$ data at $\sqrt{s}=$~3.773~GeV.
This discovery stimulated many theoretical studies and interpretations
~\cite{Soni:2020sgn,Momeni:2022gqb,Cheng:2017fkw,Huang:2021owr,Wu:2022qqx,Hsiao:2023qtk}.
Notably, since there is currently no reliable method to describe the four-quark state or molecular state picture of $a_{0}(980)$ from the first principle, 
most theoretical calculations are conducted in the two-quark scenario, such as covariant confined quark model (CCQM)~\cite{Soni:2020sgn}, AdS/QCD~\cite{Momeni:2022gqb} and light-cone sum rule (LCSR)~\cite{Cheng:2017fkw,Huang:2021owr,Wu:2022qqx}.

The $\psi(3770)$ decays predominantly to $D\bar{D}$ pairs without any additional hadrons. The excellent tracking, precision calorimetry, and the large $D\bar{D}$ threshold data sample~\cite{BESIII:2009fln} provide an unprecedented opportunity to accurately study the decay dynamics of $D^{0} \to a_{0}(980)^{-} e^+ \nu_{e}$. 
Based on a data set corresponding to an integrated luminosity of 7.93 fb$^{-1}$~\cite{Ablikim:2013ntc, BESIII:2024lbn}, this Letter reports an improved measurement of the BF for the $D^{0} \to a_{0}(980)^{-} e^+ \nu_{e}$ decay, along with the first-ever measurement of the FF for the $D \to a_{0}(980)$ transition.
Charge-conjugate modes are implied throughout this Letter.

A description of the design and performance of the BESIII detector can be found in Ref.~\cite{BESIII:2009fln}. Simulated data samples are produced with a {\sc geant4}-based~\cite{GEANT4:2002zbu} Monte Carlo~(MC) toolkit, which includes the geometric description~\cite{Huang:2022wuo} of the BESIII detector and the detector response. The simulation models the beam energy spread and initial state radiation (ISR) in the $e^+e^-$ annihilations with the generator {\sc kkmc}~\cite{Jadach:2000ir}. The inclusive MC sample includes the production of $D\bar{D}$ pairs, the non-$D\bar{D}$ decays of the $\psi(3770)$, the ISR production of the $J/\psi$ and $\psi(3686)$ states, and the continuum processes incorporated in {\sc kkmc}. All particle decays are modeled with {\sc evtgen}~\cite{Lange:2001uf, Ping:2008zz} using BFs either taken from the Particle Data Group~(PDG)~\cite{PDG}, when available, or otherwise estimated with {\sc lundcharm}~\cite{Chen:2000tv, Yang:2014vra}. Final state radiation from charged final-state particles is incorporated using {\sc photos}~\cite{Richter-Was:1992hxq}.
The contribution of the non-resonance component is found to be negligible by fitting to the data at $M_{\eta\pi^-}$ sideband region~($M_{\eta\pi^-} > 1.3~\mathrm{GeV}/c^2$).
Thus, we assume that the S-wave component of the $\eta\pi$ system solely receives contribution from the $a_{0}(980)$. The $D^{0} \to \eta \pi^{-} e^{+} \nu_{e}$ decay can be simulated according to prior measurements on SL $D$ decays ~\cite{BESIII:2015hty,BESIII:2018qmf}.
The detailed study can be found in Ref.~\cite{supplement}.
The $a_{0}(980)$ line shape is modeled by the Flatt\'e formula with its parameters fixed to the BESIII measurement~\cite{BESIII:2016tqo}.

The double-tag~(DT) method~\cite{MARK-III:1985hbd} provides high-purity samples for measuring the absolute BFs of $D$ meson decays, without the need of knowing the integrated luminosity and the $D\bar{D}$ production cross section. The $\bar{D}^0$ mesons are first reconstructed in one of the three decay modes $ K^+ \pi^-$, $ K^+ \pi^- \pi^0$, and $ K^+ \pi^- \pi^+ \pi^-$, referred to as single-tag~(ST) events. After a $\bar{D}^0$ meson is found, the $D^{0} \to \eta \pi^{-} e^{+} \nu_{e}$ candidate is searched for on the recoil side, where the presence of the undetected $\nu_{e}$ is inferred from the missing momentum.
An event in which a signal $D^{0} \to \eta \pi^{-} e^{+} \nu_{e}$ decay and an ST $\bar{D}^0$ is simultaneously found is referred to as a DT event.

The BF of the signal decay is determined by 
\begin{equation}
  \mathcal{B}=\frac{N_{\rm DT}}{\sum_{\alpha}({N^{\alpha}_{\rm ST}}\epsilon^{\alpha}_{\rm DT}/\epsilon^{\alpha}_{\rm ST})\mathcal{B}_{\rm sub}}\,,
  \label{eq:bf}
\end{equation}
where $\alpha$ denotes the tag mode, $N^{\alpha}_{\rm ST}$ is the ST yield for tag mode $\alpha$, $N_{\rm DT}$ is the sum of the DT yields from all ST modes, while $\epsilon^{\alpha}_{\rm ST}$ and $\epsilon^{\alpha}_{\rm DT}$ refer to the corresponding efficiencies as detemined from MC simulation. $\mathcal{B}_{\rm sub}$ represent the BFs of all possible intermediate particles.

In isolating the ST samples, the selection criteria of $K^{\pm}$, $\pi^{\pm}$, and $\pi^0$ candidates are the same as Ref.~\cite{BESIII:2023exq}. 
The tagged $\bar D^0$ mesons are selected using the energy difference $\Delta E \equiv  E_{\bar D^0}-E_{\rm beam}$ and the beam-constrained mass $M_{\rm BC} \equiv \sqrt{E^{2}_{\rm beam}/c^{4}-|\vec{p}_{\bar D^0}|^{2}/c^{2}}$, where $E_{\rm beam}$ is the beam energy, and $\vec{p}_{\bar D^0}$ and $E_{\bar D^0}$ are the  measured momentum and energy of the $\bar D^0$ candidate in the $e^+e^-$ rest frame, respectively.
In the case of multiple candidates in an event, the one with the minimum $|\Delta{E}|$ is chosen. The ST yields and efficiencies are determined by fitting the $M_{\rm BC}$ distributions of the accepted candidates in data and inclusive MC sample, respectively. More details about the ST selection can be found in Ref.~\cite{BESIII:2023exq}. The ST yields in the data and the corresponding ST efficiencies are listed in Table~\ref{tab: ST and DT efficiency}. Summing over all three tag modes, the total ST yield $N_{\rm ST}^{\rm tot}$ is $(6306.7\pm2.9)\times10^3$ within the $M_{\rm BC}$ signal region, ranging from 1.859 to 1.873~$\textrm{GeV}/c^{2}$.

\begin{table}[htbp]
\caption{The energy difference~($\Delta$E) windows, ST yields in data~($N_\textrm{ST}^{\alpha}$), ST efficiencies~($\epsilon_\textrm{ST}^{\alpha}$), DT efficiencies~($\epsilon_\textrm{DT}^{\alpha}$), with statistical uncertainties, for each tag mode $\alpha$. These efficiencies do not include the BFs of all possible intermediate particles.}
	 \label{tab: ST and DT efficiency}
     \resizebox{1.0\linewidth}{!}{
     \begin{tabular} {l c c c c}  
         \hline \hline 
		 Tag mode & $\Delta$E~(MeV) & $N_\textrm{ST}^{\alpha}~(\times 10^{3})$  & $\epsilon_\textrm{ST}^{\alpha}(\%)$ & $\epsilon_\textrm{DT}^{\alpha}(\%)$ \\
         \hline    
$K^+ \pi^-$ & $[-27, 27]$ & $1449.3 \pm 1.3$ & $65.34 \pm 0.01$ & $17.20 \pm 0.02$ \\$K^+ \pi^- \pi^0$ & $[-62, 49]$ & $2913.2 \pm 2.0$  &  $35.59 \pm 0.01$ & $8.88 \pm 0.01$ \\$K^+ \pi^- \pi^+ \pi^-$ & $[-26, 24]$ & $1944.2 \pm 1.6$ &  $40.83 \pm 0.01$ & $8.81 \pm 0.01$ \\
         \hline \hline 
     \end{tabular}
     }
 \end{table}

Once a ST $\bar{D}^0$ decay has been identified, a search is made for a $D^0$ signal decay in the same event.
Candidate $\eta,\pi^{-}$ and $e^{+}$ are selected with the remaining charged tracks and photon candidates. 
The selection criteria for $\pi^{\pm}$, $e^{\pm}$, and photon candidates are the same as in Refs.~\cite{BESIII:2023exq,BESIII:2023htx}. The $\eta$ candidates are reconstructed through the decay $\eta\to\gamma\gamma$, requiring the invariant mass~($M_{\gamma\gamma}$) to lie within the range (0.505, 0.575)~GeV/$c^2$. 
A  one-constraint kinematic fit is performed, constraining $M_{\gamma\gamma}$ to the known $\eta$ mass~\cite{PDG}, and the combination giving the minimum $\chi_{\rm 1C}^{2}$ is kept for further analysis.
Furthermore, DT candidates are rejected if they contain any extra charged tracks~($N_{\rm extra}^{\rm char}$) or $\pi^0$ reconstructed with unused photons~($N_{\rm extra}^{\pi^0}$). This $\pi^0$ veto effectively suppresses the backgrounds arising from $D^0 \to \pi^- \pi^0 e^+ \nu_{e}$ and $D^0 \to K^{*}(892)^- e^+ \nu_{e}$ with the  subsequent decay $K^{*}(892)^- \to K_{S}^0(\to \pi^0 \pi^0) \pi^-$. An additional background comes from $D^0 \to K^{*}(892)^- e^+ \nu_{e}$ decays, followed by $K^{*}(892)^- \to K_{L}^0 \pi^-$, where the EMC shower induced by the $K_{L}^0$ mimics the higher-energy photon of the $\eta$ candidate. To suppress these events, the lateral moment~\cite{Drescher:1984rt} of EMC showers, which peaks around 0.15 for real photons but varies from 0 to 0.85 for $K_{L}^0$ candidates, is required to lie within (0, 0.35) for the higher-energy photon from the $\eta$ candidate. This requirement suppresses about 78\% of the $K_{L}^0$ background while retaining 93\% of the signal.

The energy and momentum of the undetectable neutrino in the $e^{+}e^{-}$ CM frame are derived as $E_{\rm miss} \equiv E_{\rm beam} - \sum_{i}E_{i}$ and $\vec{p}_{\rm miss} \equiv -(\vec{p}_{\rm tag} + \sum_{i}\vec{p}_{i})$, respectively, where $E_{i}$ and $\vec{p}_{i}$ are the energy and momentum of the $\eta,\pi^{-}$, and $e^{+}$ of the signal side. 
We calculate $\vec{p}_{\rm tag}=\hat{p}_{\rm tag}\sqrt{E_{\rm beam}^{2}-m_{D^0}^{2}c^4}$, where $\hat{p}_{\rm tag}$ is the unit vector in the momentum direction of the ST $D^0$ and $m_{D^0}$ is the known $D^0$ mass~\cite{PDG}.
The signal candidates are expected to peak around zero in the $U_{\rm miss} \equiv E_{\rm miss} - c|\vec{p}_{\rm miss}|$ distribution and near the nominal mass of $a_0(980)$ in the $\eta\pi^-$ mass spectrum~($M_{\eta\pi}$).

\begin{figure}[htbp]
	\begin{center}
    \includegraphics[width = 0.5\textwidth]{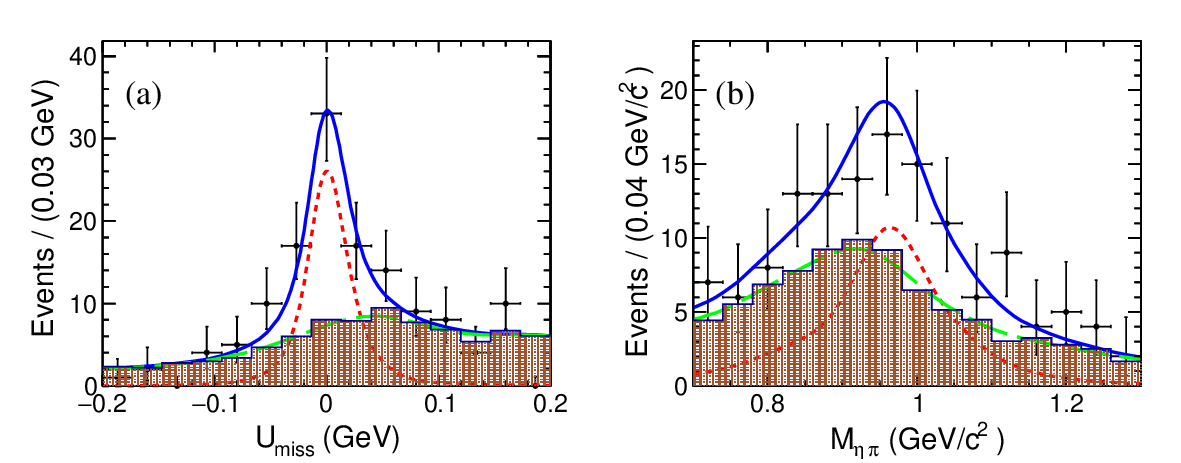}
	\end{center}
    \vspace{-0.5cm}
	\caption{
Projections of the 2D fit on $U_{\textrm{miss}}$~(a) and $M_{\eta\pi^-}$~(b) for $D^{0} \to a_{0}(980)^{-} e^+ \nu_{e}$.  The points with error bars are data. The (blue) solid curves are the overall fits, the (red) short-dashed lines show the fitted signal shape; the (green) long-dashed lines denote the background shape. 
	}
	\label{fig:Fitresult}
\end{figure}

To obtain the signal yields, we perform two-dimensional~(2D) unbinned maximum likelihood fits to the $U_{\rm miss}$ versus $M_{\eta\pi}$ distributions.
The signal shape in the $U_{\rm miss}$ distribution is described by the MC simulation and that in the $M_{\eta\pi}$ distribution is modeled with a standard Flatt\'e formula~\cite{Flatte:1976xu} for the $a_0(980)$ signal, using 0.990~$\textrm{GeV}/c^{2}$ for the $a_0(980)$ mass ($m_{0}$). 
Here we assume that $a_0(980)$ decays only to $\eta \pi$ and $K\bar{K}$ final states, with the corresponding coupling constants fixed at $g_{\eta \pi}^2 = 0.341\,(\textrm{GeV}/c^{2})^2$ and $g_{K \bar{K}}^2=0.304\,(\textrm{GeV}/c^{2})^2$, respectively~\cite{BESIII:2016tqo}.
The background is described by the shape found in the inclusive MC sample. The 2D probability density functions of all these components are constructed by the product of the $U_{\rm miss}$ and $M_{\eta\pi}$ 
functions. Studies performed with the inclusive MC simulation shows that there is a  negligible correlation between these two observables.
Projections of the 2D fits are shown in Fig.~\ref{fig:Fitresult}.
The bump around 0.9~GeV of the background shape in the $M_{\eta \pi}$ mass spectrum predominantly originates from the decay $D^0 \to K^{*-} e^+ \nu_{e}$. The impact on BF is discussed in the next paragraph.
Without considering the non-resonance contribution, the measured signal yield is $N_{\rm DT}=51.8\pm10.0$ with a statistical significance of $6.9\sigma$.
The statistical significance is given by $\sqrt{-2 \ln(\mathcal{L}_{0}/\mathcal{L})}$,  where $\mathcal{L}$ is the maximum likelihood value with the signal yield as a free parameter, and $\mathcal{L}_{0}$ is that with the signal yield fixed to zero.
The corresponding DT efficiencies are presented in Table~\ref{tab: ST and DT efficiency}.

The systematic uncertainties in the BF measurement are summarized in Table~I of Ref.~\cite{supplement} and discussed below.
The uncertainty in the ST $\bar{D}^0$ yield reflects the uncertainty in the fit to the $M_{\rm BC}$ distributions and is assessed by varying the signal and background shapes.
The uncertainties in the tracking or PID efficiencies of $\pi^{-}$ and $e^+$ are studied with control samples of DT-hadronic events and $e^{+}e^{-} \to \gamma e^{+}e^{-}$, respectively.
Due to the consistency in the overall reconstruction efficiency between $\eta$ and $\pi^0$, the uncertainty associated with the $\eta$ reconstruction and $N_{\rm extra}^{\pi^0}$ requirement are both assigned by studying a control sample of $D^0 \to K^- \pi^+ \pi^0$ decays.
The efficiency of the lateral moment requirement for photons is studied in different energy and polar-angle bins using a control sample of $e^{+}e^{-} \to \gamma e^{+}e^{-}$ events. 
The efficiency correction factors for these sources are determined by measuring the differences in efficiency between the MC and data control samples, which are weighted based on the distributions of several common kinematic variables, including energy, momentum, and polar angle, of the signal MC sample. The resulting systematic uncertainties are then propagated from the statistical uncertainties of the control samples.
The uncertainty arising from the signal model is estimated by measuring the BF difference when changing the line shape parameters~($m_{0}, g_{\eta \pi}^{2}, g_{K\bar{K}}^2$) of the $a_0(980)$ measured by BESIII collaboration~\cite{BESIII:2016tqo} to the one measured by CLEO-c collaboration~\cite{CLEO:2011upl} in the signal MC generator.
The uncertainty of the background shape is assessed by altering the baseline MC background shape with two methods, and subsequently combining the uncertainties derived from these alterations to represent the overall uncertainty.
Firstly, to account for the imperfect agreement between data and MC simulation for $K_{L}^0$ reconstruction, we increase the yield of the major background process $D^0 \to K^{*}(892)^- e^+ \nu_{e}$ with $K^{*}(892)^- \to K_{L}^0 \pi^-$ by 30\%, based on the analysis of control samples $D^0 \to K_{L}^0 \pi^+ \pi^-$, $D^0 \to K_{L}^0 \pi^+ \pi^- \pi^0$ and $D^+ \to K_{L}^0 \pi^+ \pi^0$ decays.
Secondly, the $e^{+}e^{-} \to q\bar{q}$ background yields are changed by $\pm 1\sigma$ of the measured cross section~\cite{Ablikim:2007zz}.
In order to assign a systematic uncertainty associated with the modelling of the  resolution of $U_{\rm miss}$, the fit is reperformed with the signal shape convolved with a Gaussian function. The width of the Gaussian distribution, which represents the resolution difference between experimental data and MC simulation, is fixed based on the estimation derived from the control sample of the decay process $D^{0} \to \pi^{0} \pi^{-} e^{+} \nu_{e}$. Changes in the signal yields are assigned to be the corresponding uncertainties.
The uncertainty associated with  the assumed BF of the $\eta$ decay is taken from Ref.~\cite{PDG}.
The effect of the limited MC sample size is also included as a systematic effect.
Assuming that all sources are independent, the total systematic uncertainty on the BF is determined to be 5.2\%, by adding them in quadrature. 


To extract the hadronic FF of the $D\to a_{0}(980)$ transition, the signal candidate events are divided into three intervals of $q^2$~(the four-momentum transfer squared of the $e^{+} \nu_{e}$ system) and a $\chi^2$ fit is used to determine the partial decay rates. Considering the correlations of the measured partial decay rates~($\Delta\Gamma^{i}_{\rm mea}$) among different $q^2$ intervals, the $\chi^2$ is given by
\begin{equation}
    \chi^{2} = \sum_{ij} (\Delta\Gamma^{i}_{\rm mea} - \Delta\Gamma^{i}_{\rm exp}){[C^{-1}]}_{ij}(\Delta\Gamma^{j}_{\rm mea} - \Delta\Gamma^{j}_{\rm exp}),
    \label{eq:chi2}
\end{equation}
where $\Delta\Gamma_{\rm mea}^{i}$ and $\Delta\Gamma_{\rm exp}^{i}$ represent the measured and theoretically expected partial decay rates, respectively, in the $i$-th $q^2$ interval. Here, the indices $i$ and $j$ denote different $q^2$ intervals.
The elements of the inverse covariance matrix $C^{-1}$, denoted as $[C^{-1}]_{ij}$, account for the correlations between the different $q^2$ bins.

In the limit of massless leptons, the differential decay rate is dominated by the term involving $f_{+}(q^2)$~\cite{Cheng:2017fkw}.
Hence, the $\Delta\Gamma_{\rm exp}^{i}$ are calculated by integrating the following double differential decay rate~\cite{Wang:2016wpc,Achasov:2020qfx}:
\begin{equation}
\begin{split}
&\frac{{\rm d}^{2}\Gamma(D^{0} \to a_{0}(980)^{-} e^+ \nu_{e})}{{\rm d}s{\rm d}q^2}  \\ &=\frac{G^2_F|V_{cd}|^{2}}{192\pi^{4}m_{D^0}^{3}} \lambda^{3/2}(m_{D^0}^{2},s,q^2)|f_{+}^{a_0}(q^2)|^2P(s),
\end{split}
\label{eq:DDR}
\end{equation}
where $s$ is the square of $M_{\eta \pi}$, $G_F$ is the Fermi constant~\cite{PDG}, $V_{cd}$ is the Cabibbo-Kobayashi-Maskawa matrix element, $m_{D^0}$ is the known $D^0$ mass~\cite{PDG}, $\lambda(x,y,z)=x^2+y^2+z^2-2xy-2xz-2yz$, and $P(s)$ is based on the relativistic Flatt\'e formula~\cite{Flatte:1976xu,CLEO:2011upl}:
\begin{equation}
P(s)=\frac{g_{\eta \pi}^2\rho_{\eta\pi}}{|m^2_{0}-s-i(g_{\eta \pi}^2\rho_{\eta \pi}+g_{K \bar{K}}^2\rho_{K \bar{K}})|^2},
\label{eq:flatte}
\end{equation}
where 
$\rho_{\eta \pi}$ and $\rho_{K\bar{K}}$ are individual phase-space factors. 
The product of the FF and $|V_{cd}|$ is extracted in three intervals of $q^2$. 
The FF is modeled with the single-pole parameterization~\cite{Becirevic:1999kt}:
\begin{equation}
f_{+}^{a_0}(q^2)=\frac{f_{+}^{a_0}(0)}{1-\frac{q^{2}}{m_{\mathrm{pole}}^{2}}},
\label{eq:single pole}
\end{equation}
where $f_{+}^{a_0}(0)$ is the FF evaluated at $q^2=0$~GeV$^{2}$/$c^{4}$, and the pole mass $m_{\mathrm{pole}} = 2.42$~GeV/$c^2$~\cite{PDG,Achasov:2021dvt}.

The measured partial decay rate $\Delta\Gamma^{i}_{\rm mea}$ is determined by
$\Delta\Gamma^{i}_{\rm mea}= N_{\rm pro}^{i}/(\tau N_{\rm ST}^{\rm tot})$
, where $\tau$ is the $D^0$-meson lifetime~\cite{PDG} and $N_{\rm pro}^{i}$ is the signal yield produced in the $i$-th $q^2$ interval, given by $N_{\rm pro}^{i}=\sum_{j=1}^{3}[\epsilon^{-1}]_{ij}N_{\rm obs}^{j}$.
The $N_{\rm obs}^{j}$ is the observed signal yield obtained from the 2D fits on the $U_{\textrm{miss}}$ versus $M_{\eta\pi^-}$ distribution in the $j$-th $q^2$ interval, carried out in a similar manner as the one described previously for the BF measurement.
The matrix $\epsilon^{-1}$ represents the inverse of the efficiency matrix.  $\epsilon_{ij}$ is the efficiency matrix element  determined from the signal MC samples via
$\epsilon_{ij}=\sum_{k}[(1/N_{\rm ST}^{\rm tot})\times(N_{\rm rec}^{ij}/N_{\rm gen}^{j})_{k}\times(N_{\rm ST}^{k}/\epsilon_{\rm ST}^{k})]$, where $N_{\rm rec}^{ij}$ is the number of signal events reconstructed in the $i$-th $q^2$ interval while generated in the $j$-th $q^2$ interval, $N_{\rm gen}^{j}$ represents the generated number of signal events in the $j$-th $q^2$ interval, and $k$ sums over all tag modes. The details of the $q^2$ divisions, $N_{\rm obs}^{i}$, $N_{\rm pro}^{i}$, and $\Delta\Gamma^{i}_{\rm mea}$ are given in Table~\ref{tab:deltaGam_D0}.

\begin{table}[htbp]
  \centering
  \caption{The partial decay rates of $D^0 \to a_{0}(980)^{-} e^+ \nu_{e}, a_{0}(980)^{-} \to \eta \pi^{-}$ in different $q^{2}$ intervals, where the uncertainties are statistical only.}
    \begin{tabular}{c c c c c}
    \hline \hline
    $i$ & $q^{2}$(GeV$^{2}$/$c^{4}$) & $N_{\rm obs}^{i}$   &  $N_{\rm pro}^{i} $ &  $\Delta\Gamma^{i}_{\rm mea}(\rm ns^{-1})$ \\
    \hline
1 &$[0.0, 0.2]$ & $ 16.9\pm5.7$ & $ 75.6\pm27.0$ & $ 0.074\pm0.027$ \\ 2 &$[0.2, 0.4]$ & $ 15.5\pm5.0$ & $ 63.1\pm22.4$ & $ 0.062\pm0.022$ \\ 3 &$[0.4, q_{\rm max}^{2}]$ & $ 17.4\pm5.9$ & $ 67.6\pm23.8$ & $ 0.066\pm0.023$ \\
    \hline \hline
    \end{tabular}
\label{tab:deltaGam_D0}
\end{table}

The statistical and systematic covariance matrices are
constructed as $C_{ij}^{\rm stat} = (\frac{1}{\tau N_{\rm ST}^{\rm tot}})^{2}\sum_{\alpha}\epsilon_{i\alpha}^{-1}\epsilon_{j\alpha}^{-1}\sigma^{2}(N^{\alpha}_{\rm obs})$ and $C_{ij}^{\rm syst}=\delta(\Delta\Gamma^{i}_{\rm mea}) \delta(\Delta\Gamma^{j}_{\rm mea})$, respectively, where $\sigma(N^{\alpha}_{\rm obs})$ and $\delta(\Delta\Gamma^{i}_{\rm mea})$ are the statistical and systematic uncertainties in the $i$-th $q^2$ interval. The $C_{ij}^{\rm syst}$ elements are obtained by summing all the covariance matrices for all systematic uncertainties, where the systematic uncertainty of $\tau$, 0.2\%~\cite{PDG}, is included in addition to those in the BF measurement and are given in Ref.~\cite{supplement}.

The systematic uncertainty related to $\Delta\Gamma^{i}_{\rm mea}$ is estimated to be 2.4\% by following Ref.~\cite{BESIII:2015tql}. In addition, the input parameters $m_{0}$, $g_{\eta \pi}^2$ and $g_{K \bar{K}}^2$ related to $\Delta\Gamma^{i}_{\rm exp}$ are also considered by varying them within $\pm 1\sigma$ from their central values~\cite{BESIII:2016tqo}. The largest deviations of the form factor, respectively 0.2\%, 0.1\% and 0.6\%, are taken as systematic uncertainties. The quadrature sum of these uncertainties is 2.7\%, and is taken as the total systematic uncertainty.

\begin{figure}[htbp]
    \begin{center}
        \mbox{
            \put(-130, 0){
            \includegraphics[width=0.25\textwidth]{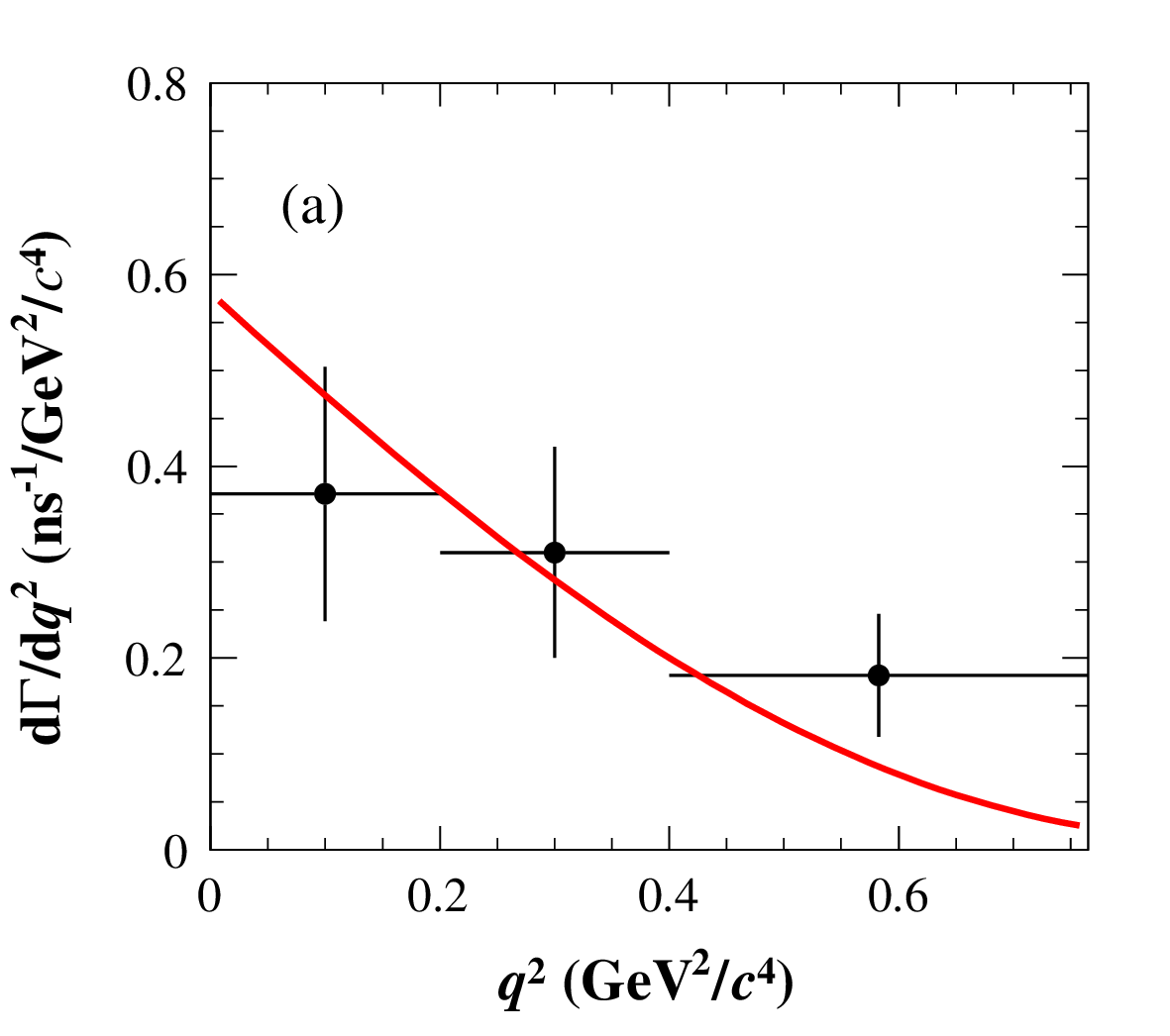}
            }
            \put(0, 0){
            \includegraphics[width=0.25\textwidth]{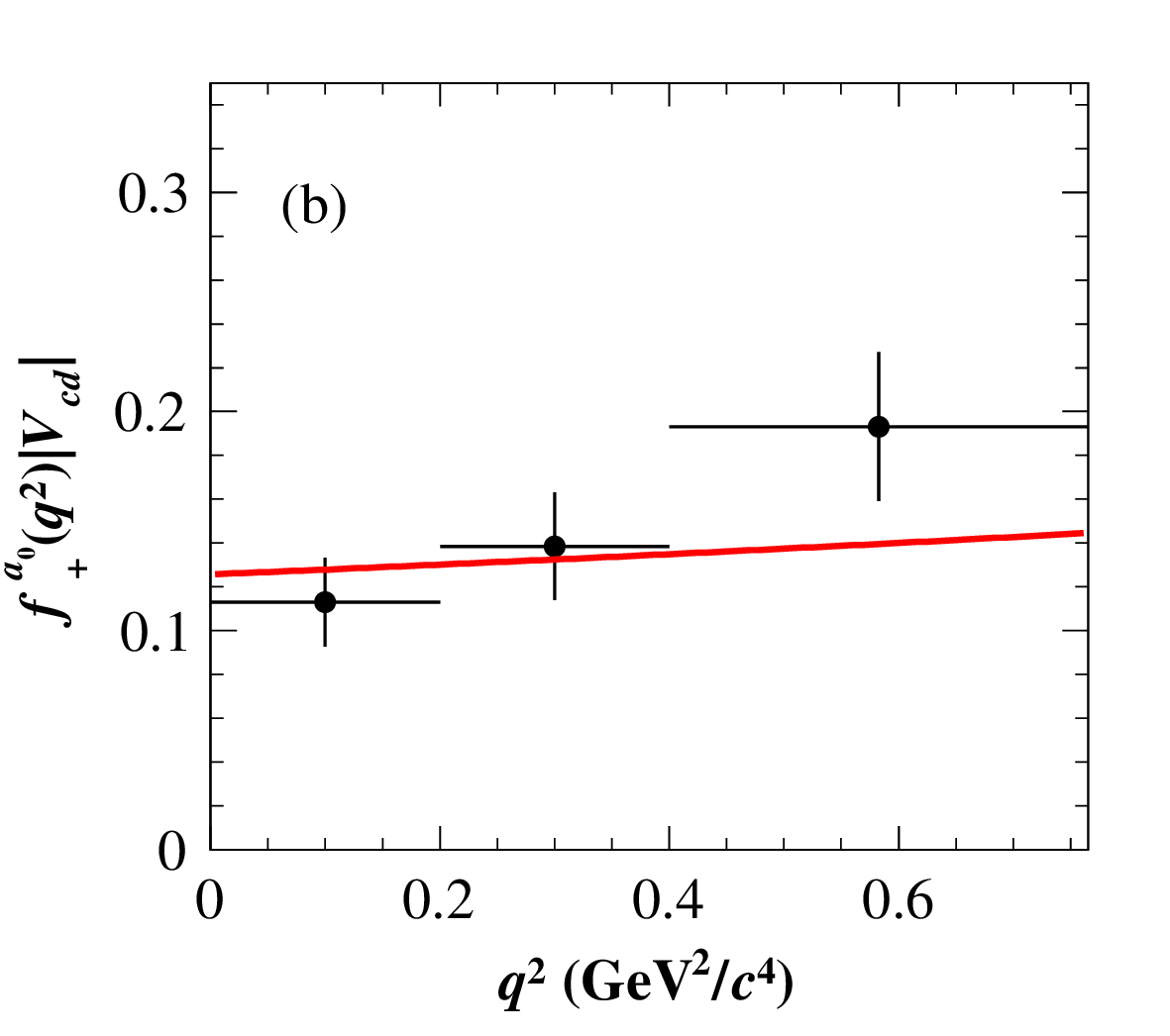}
            }
        }
    \end{center}
    \vspace{-0.5cm}    
    \caption{
Fit to the differential decay rate as function of $q^2$ (a) and projection to the FF $f^{a_{0}}_{+}(q^2)|V_{cd}|$ (b).
The points with error bars are data. The red solid line is the fit result.
    }
    \label{fig:DDR&&FF}
\end{figure}

A fit is performed for the differential decay rate measured through the process $D^{0} \to a_0(980)^{-} e^{+} \nu_{e}$ and the results are presented in Figs.~\ref{fig:DDR&&FF}~(a) and (b) where the fitted decay rate and its projection onto the hadronic FF.
The goodness-of-fit is $\chi^2/\rm NDOF = 2.2/2=1.1$, where NDOF is the number of degrees of freedom.
The product $f_{+}^{a_{0}}(0)|V_{cd}|$ is determined to be $0.126\pm0.013_{\rm stat}\pm0.003_{\rm syst}$.
The baseline FF is taken from the fit with the combined statistical and systematic covariance matrix, and the statistical uncertainty on the FF is taken from the fit with only the statistical covariance matrix. The systematic uncertainty is obtained by calculating the quadratic difference of uncertainties between these two fits.

Using $e^+e^-$ collision data corresponding to an integrated luminosity of 7.93${~\rm fb^{-1}}$ collected with the BESIII detector at $\sqrt{s} = {\rm 3.773~GeV}$, this Letter reports the absolute BF of the decay $D^0 \to a_{0}(980)^{-} e^+ \nu_{e}$ with $a_{0}(980)^{-} \to \eta \pi^-$ to be $(0.86\pm0.17_{\text{ stat}}\pm0.05_{\text{ syst}})\times 10^{-4}$, which is 1.2 times more precise than the previous measurement~\cite{BESIII:2018sjg}.
The improved precision amplifies the difference of calculations in the two- and four-quark scenarios.
Our result exhibits some tension with theoretical predictions with the two-quark model, as shown in Table~\ref{tab:result comparison} and favor, according to Ref.~\cite{Hsiao:2023qtk}, the prediction based on the four-quark picture.

Furthermore, we have investigated the dynamics of the decay $D^0 \to a_{0}(980)^{-} e^+ \nu_{e}, a_{0}(980)^{-} \to \eta \pi^-$ and measured the product $f_{+}^{a_{0}}(0)|V_{cd}|=0.126\pm0.013_{\rm stat}\pm0.003_{\rm syst}$ for the first time by parametrizing the FF using a single-pole model.
With the value of $|V_{cd}|=0.22486\pm0.00067$ from the standard model global fit~\cite{PDG} as input, we determine $f_{+}^{a_{0}}(0)=0.559\pm0.056_{\rm stat}\pm0.013_{\rm syst}$.
Table~\ref{tab:result comparison} shows that these results are consistent with the CCQM~\cite{Soni:2020sgn}, SU(3) flavor symmetry~\cite{Hsiao:2023qtk}
within 3$\sigma$,
but disfavor the Ads/QCD~\cite{Momeni:2022gqb} and LCSR calculations~\cite{Cheng:2017fkw,Huang:2021owr,Wu:2022qqx} by more than 3$\sigma$. 
This suggests the possibility that more advanced models, instead of the traditional two-quark, are needed to describe the structures of scalar mesons.

The measured BF of $D^0 \to a_{0}(980)^{-} e^+ \nu_{e}$ and the FF of $D$  to $a_{0}(980)$ transition will serve as an important input for understanding the intricate internal structure of the light scalar mesons~\cite{Wang:2009azc,Hsiao:2023qtk}, and the nonperturbative dynamics of charm meson decays.
The measurement of  $f_{+}^{a_{0}}(0)$ has additional importance  in the information it gives on the   direct hadronization of the $D$ to $a_{0}(980)$ transition. This information is essential input for performing an accurate calculation of the ratio of external $W$-emission and $W$-annihilation contributions in the quasi-two-body $D\to SP$ sector when estimating the contributions from final-state interactions~\cite{Cheng:2022vbw}.
In the near future, an investigation based on the full 20 fb$^{-1}$ of data now available at BESIII detector will enable a more precise exploration of the production of light scalar states in the SL decays of $D$ mesons~\cite{BESIII:2020nme,Ke:2023qzc,Li:2021iwf}.

\begin{table*}[htbp]
    \caption{Comparisons of $f_{+}^{a_{0}}(0)$ and $\mathcal{B}(D^{0} \to a_{0}(980)^{-} e^+ \nu_{e}, a_{0}(980)^{-} \to \eta \pi^-)$ between previous measurements, theoretical predictions, and the present experimental results reported in this work.
    Assuming that the  $a_{0}(980)$ width is saturated by the $K\Bar{K}$ and $\eta \pi$ modes, $\mathcal{B}(a_{0}(980) \to \eta \pi)$ can be inferred from the PDG average value of $\Gamma(a_0(980) \to K\Bar{K})/\Gamma(a_0(980) \to \eta \pi)=0.172\pm0.019$~\cite{PDG}.
    The narrow width approximation is further applied to estimate the $\mathcal{B}$ $\mathit{l}$ for uncomputed cases~(denoted by the superscript *).
    }
    \label{tab:result comparison}
    \begin{center}
    \renewcommand{\arraystretch}{1.16}
    \begin{tabular*} {0.9\linewidth} {l c c}
                \hline \hline 
                Theory or experiment & $f_{+}^{a_{0}}(0)$ & $\mathcal{B}(D^{0} \to a_{0}(980)^{-} e^+ \nu_{e}, a_{0}(980)^{-} \to \eta \pi^-)$~( $\times 10^{-4}$)\\ 
                \hline CCQM~\cite{Soni:2020sgn} & $0.55_{-0.02}^{+0.02}$ & $1.43\pm0.13^*$\\
                Ads/QCD~\cite{Momeni:2022gqb} & $0.72\pm0.09$ & $2.08\pm0.26^*$\\
                LCSR 2017~\cite{Cheng:2017fkw} & $1.75_{-0.27}^{+0.26}$ & ${3.48_{-1.04}^{+1.17}}^*$\\
                LCSR 2021~\cite{Huang:2021owr} & $0.85_{-0.11}^{+0.10}$ & 1.15\\
                LCSR 2023~\cite{Wu:2022qqx} & $1.058_{-0.035}^{+0.068}$ & $1.330_{-0.134}^{+0.216}$\\
                SU(3) flavor symmetry~\cite{Hsiao:2023qtk} & $0.46\pm0.06$ & $\dots$\\
                \hline
                BESIII 2018~\cite{BESIII:2018sjg} & $\dots$ & ${1.33_{-0.29}^{+0.33}}_{\rm stat} \pm 0.09_{\text{syst}}$\\
                This work & $0.559\pm0.056_{\rm stat}\pm0.013_{\rm syst}$ & $0.86\pm0.17_{\text{stat}}\pm0.05_{\text{syst}}$  \\
        \hline \hline
    \end{tabular*}
    \end{center}
\end{table*}


The BESIII Collaboration thanks the staff of BEPCII and the IHEP computing center for their strong support. This work is supported in part by National Key R\&D Program of China under Contracts Nos. 2023YFA1606000, 2020YFA0406400, 2020YFA0406300; National Natural Science Foundation of China (NSFC) under Contracts Nos. 11635010, 11735014, 11935015, 11935016, 11935018, 12025502, 12035009, 12035013, 12061131003, 12192260, 12192261, 12192262, 12192263, 12192264, 12192265, 12221005, 12225509, 12235017, 12361141819; the Chinese Academy of Sciences (CAS) Large-Scale Scientific Facility Program; the CAS Center for Excellence in Particle Physics (CCEPP); Joint Large-Scale Scientific Facility Funds of the NSFC and CAS under Contract No. U2032104, U1832207; 100 Talents Program of CAS; the Excellent Youth Foundation of Henan Scientific Committee under Contract No. 242300421044; the Fundamental Research Funds for the Central Universities under Contract No. 020400/531118010467; The Institute of Nuclear and Particle Physics (INPAC) and Shanghai Key Laboratory for Particle Physics and Cosmology; German Research Foundation DFG under Contracts Nos. FOR5327, GRK 2149; Istituto Nazionale di Fisica Nucleare, Italy; Knut and Alice Wallenberg Foundation under Contracts Nos. 2021.0174, 2021.0299; Ministry of Development of Turkey under Contract No. DPT2006K-120470; National Research Foundation of Korea under Contract No. NRF-2022R1A2C1092335; National Science and Technology fund of Mongolia; National Science Research and Innovation Fund (NSRF) via the Program Management Unit for Human Resources \& Institutional Development, Research and Innovation of Thailand under Contracts Nos. B16F640076, B50G670107; Polish National Science Centre under Contract No. 2019/35/O/ST2/02907; Swedish Research Council under Contract No. 2019.04595; The Swedish Foundation for International Cooperation in Research and Higher Education under Contract No. CH2018-7756; U.S. Department of Energy under Contract No. DE-FG02-05ER41374.

\bibliography{References.bib}

\end{document}


\normalsize
\parskip=5pt plus 1pt minus 1pt
\title{
\boldmath Supplemental Material: Study of the light scalar meson $a_{0}(980)$ through the decay $D^{0} \to  a_{0}(980)^{-} e^+ \nu_{e}$ with $a_{0}(980)^- \to \eta \pi^-$}
\date{\today}
\maketitle
Table~\ref{tab:BF systematic} summarizes the sources of the systematic uncertainties in the BF measurement of the decay $D^0 \to a_{0}(980)^{-} e^+ \nu_{e}$ with $a_{0}(980)^{-} \to \eta \pi^-$.

Table~\ref{tab:cov_matrix} summarizes the covariance matrices with statistical and systematic contributions.

\begin{table}[htbp]
    \caption{Relative systematic uncertainties in the BF measurement.}
    \label{tab:BF systematic}
    \begin{center}
    \begin{tabular} {l c}
        \hline \hline
        Source  &  Uncertainty $(\%)$ \\
        \hline
        $N_{\rm ST}^{\rm tot}$  & 0.1  \\
        $\pi^{\pm}$, $e^{\pm}$ tracking & 0.3 \\
        $\pi^{\pm}$, $e^{\pm}$ PID & 0.3 \\
        $\eta$ reconstruction & 1.6  \\
        $N_{\rm extra}^{\pi^0}$ requirement  & 0.7  \\
        Lateral moment cut & 1.1 \\
        Signal model  & 3.5  \\
        Background shape & 1.3 \\ 
        $U_{\rm miss}$ resolution & 2.9\\
        $\mathcal{B}_{\eta \to \gamma \gamma}$   & 0.5   \\
        MC sample size       & 0.1  \\
        \hline
        Total             & 5.2   \\
        \hline \hline
    \end{tabular}
    \end{center}
\end{table}

\begin{table}[htbp]
\centering
\caption{Statistical and systematic covariance matrices for the decay $D^0 \to a_{0}(980)^{-} e^+ \nu_{e}$ with $a_{0}(980)^{-} \to \eta \pi^{-}$. 
The total covariance matrix is $C_{ij}=C_{ij}^{\rm stat.}+C_{ij}^{\rm syst.}$.}
\label{tab:cov_matrix}
\begin{tabular}{c c c c}
    \hline \hline
    $C_{ij}^{\rm stat.}$   &   1   &   2   &   3   \\
    \hline
1&0.000705&-0.000055&0.000002\\2&-0.000055&0.000484&-0.000048\\3&0.000002&-0.000048&0.000547\\
    \hline \\ \\ \hline
    $C_{ij}^{\rm syst.}$   &   1   &   2   &   3   \\ \hline
    1&0.000032&0.000001&0.000002\\2&0.000001&0.000008&-0.000001\\3&0.000002&-0.000001&0.000050\\
    \hline \\ \\ \hline
    $C_{ij}$   &   1   &   2   &   3   \\ \hline
    1&0.000737&-0.000054&0.000004\\2&-0.000054&0.000492&-0.000049\\3&0.000004&-0.000049&0.000597\\
    \hline \hline
\end{tabular}
\end{table}